\documentclass[epj,nopacs]{svjour}

\usepackage{graphicx}
\usepackage{subfigure}
\usepackage[dvipdfm,bookmarks,bookmarksnumbered]{hyperref}
\usepackage[tbtags]{amsmath}

\begin{document}

\title{Rip/singularity free cosmology models with bulk viscosity}
\author{
Xin-he Meng\inst{1,2} \thanks{\email{xhm@nankai.edu.cn}}
\and Zhi-yuan Ma \inst{1} \thanks{\email{mazhiyuan@mail.nankai.edu.cn}}
}
\institute{Department of physics, Nankai University, Tianjin 300071, China
\and Kavli Institute of Theoretical Physics China, CAS, Beijing 100190, China}
\date{Received: date / Revised version: date}
\abstract{ In this paper we present two concrete models of
non-perfect fluid with bulk viscosity to interpret the observed
cosmic accelerating expansion phenomena, avoiding the introduction
of exotic dark energy. The first model we inspect has a viscosity of
the form $\zeta=\zeta_{0}+(\zeta_{1}-\zeta_{2}q)H$ by taking into
account of the decelerating parameter q, and the other model is of
the form $\zeta=\zeta_0+\zeta_1H+\zeta_2H^2$. We give out the exact
solutions of such models and further constrain them with the latest
Union2 data as well as the currently observed Hubble-parameter
dataset (OHD), then we discuss the fate of universe evolution in
these models, which confronts neither future singularity nor
little/pseudo rip. From the resulting curves by best fittings we
find a much more flexible evolution processing due to the presence
of viscosity while being consistent with the observational data in
the region of data fitting. With the bulk viscosity considered, a
more realistic universe scenario is characterized comparable with
the $\Lambda$CDM model but without introducing the mysterious dark
energy. } \maketitle
\section{Introduction}
The independent discovery respectively in 1998 and 1999 indicates
that the current universe is in accelerating
expansion~\cite{aexpansion}. To accommodate this surprisingly exotic
phenomenon, dark energy, varieties of models are proposed. The basic
idea of dark energy comes up in the context of supposing the general
theory of relativity works precisely well in cosmological scale, a
perfect fluid with effectively large enough negative pressure is
required to speed the universe expansion up. According to the
Wilkinson Microwave Anisotropy Probe (WMAP) 7-year dataset
analysis~\cite{wmap7} it makes up about $72.8\%$ of the universe's
contents. In the past decade, many attempts are made to understand
the accelerating mechanism of dark energy, the most mysterious
component of the universe hitherto as envisioned. The $\Lambda$CDM
is such a model and receives great attention because it is well
consistent with the observational data although suffering from some
serious fundamental physics problems~\cite{lcdm}. Another popular
theory of dark energy is referred as
quintessence~\cite{quintessence}. However, an alternatively
instructive idea is that the general theory of relativity may fail
in large scale, therefore modification of gravity theory should be
introduced to drive an accelerating phase of universe expansion. The
$f(R)$ gravity, for example, which generalizes the Einstein-Hilbert
action by invoking an arbitrary function of Ricci scalar, is of this
class~\cite{fR}. More details and references are available in the
recent reviews~\cite{reviews}.

In the context of perfect fluid, models with different equation of
state are extensively studied (see Ref.~\cite{fluid models}), such as Chaplygin gas, generalized
Chaplygin gas, inhomogeneous equation of state, and barotropic fluid
dark energy, {\it etc}. However, the perfect fluid
assumption is assertive since it suggests no dissipation, which
actually exists widely and intuitively plays a critical role in the
universe evolution, especially in the early hot stages. To be more
realistic, models of imperfect fluid are invoked by introducing
viscosity into the investigation. The simplest theory of the kind is
constant bulk viscosity, regarded as equivalence to constant dark
energy with cold dark matter model. A widely investigated case is
that bulk viscosity with the form as a linear function of the
Hubble parameter, which is proved to be well consistent with the
observed late-time acceleration and can recover both the matter and
dark energy dominant eras. Another case is that bulk viscosity
behaves as a parameterized power-law form with respect to the matter
density, which can be shown to be similar to the Chaplygin gas
model~\cite{fluid models2}. Furthermore, viscosity term contains
red-shift and higher derivatives of scale factor is also considered
as a more general theory~\cite{douxu}. On the other hand, as subsequence of viscosity and dissipation, turbulence is considered as a cosmic component in Ref.~\cite{turbu}. In this paper we focus on the bulk viscosity and propose two
such new models, which can be dealt with analytically and performed
well when comparing with observational data.

It is well known that significant number of dark energy models
suffer from the finite-time singularity problem. 
The classification of the (four) finite-time future singularities
has been proposed by S.~Nojiri {\it et al.} (2005)~\cite{singupro}, while the big rip (Type I) and sudden rip (Type II) are discussed previously in Ref.~\cite{singularity} and~\cite{futurerip}, respectively. Recently, a novel concept
has been proposed in Ref.~\cite{littlerip}, the so-called ``Little
Rip", in which for some kinds of models the non-viscous dark energy
density increases with equation of state parameter $\omega<-1$, but
that $\omega\rightarrow -1$ asymptotically is required to avoid the
future singularity, thereby leading to a scenario that bound
structure is disassembled (see Ref.~\cite{littleripmodel} for more concrete models of this class). Likewise, ``Pseudo Rip", which is
introduced as another case of the rips, occurs when an upper bound
$F_{inert}$ where $F$ refers to the inertial force is confronted
during the increasing of dark energy density with scale factor.
Several models have been studied as this type~\cite{pseudorip}. Further more, models reconstructed to realize or avoid such rip/singularity in scheme of modified gravity are also studied recently in Ref.~\cite{reconstructionrip}. The
presence of rips causes destruction of cosmos structure, therefore
it is an important aspect for a cosmological viable model building
that a natural scenario can be yielded to cure such
problem~\cite{singufree}. In the scheme of viscous cosmology, singularity as a vital feature attracts great attention (see most of the models in Ref.~\cite{fluid models}\cite{fluid models2}). Recently, the rips confronted in viscosity models are also studied in Ref.~\cite{viscous rip}. Generally, it implies that the presence
of particular viscosity could mitigate the singularness of cosmos
quantities in some extends.

In this letter, we propose and investigate two different viscosity
models, both of which are constructed under the assumption that our
universe is bound more tightly therefore the bulk viscosity varies
with respect to not only the conventional momentum term $H$, but
also the accelerating status ($\dot{H}$, model I) and total energy
($H^2$, model II). Arrangement for this article is as follows. In the
next section, we derive the explicit solution of the models,
respectively, and calculate analytically the corresponding cosmic
quantities. In Section~\ref{fit}, we briefly review the statistics
analysis employed for constraining model parameters by SNe Type Ia
date and OHD dataset. In Section~\ref{result} figures and tables
are listed as our results, in which the comparison with the
$\Lambda$CDM model is also performed. In Section~\ref{conclusion},
as our conclusion, we discuss the result of best-fitting and the
fate of universe evolution accordingly by means of the singularity
and rip analysis, as well as the advantages and disadvantages of
such viscosity models. For the benefit to the related astrophysics
and cosmology community we list the latest 18 OHD in the appendix.

\section{Cosmology Models of Imperfect Fluid}\label{solution}
In the scheme of imperfect fluid, the energy stress tensor reads,
\begin{equation}\label{equ:Tmunu}
T_{\mu\nu}=\rho U_{\mu}U_{\nu}+(p-\theta\zeta)h_{\mu\nu}-2\eta\sigma_{\mu\nu}+Q_{\mu}U_{\nu}+Q_{\nu}U_{\mu},
\end{equation}
where $\rho$ is the mass density, $p$ the isotropic pressure, $U^{\mu}=(1,0,0,0)$ the four-velocity of the cosmic fluid in comoving coordinates, $h_{\mu\nu}=g_{\mu\nu}+U_{\mu}U_{\nu}$ the projection tensor, $\theta\equiv\theta^{\mu}_{\mu}=h^{\alpha}_{\mu}h^{\beta}_{\mu}U_{(\alpha;\beta)}=U^{\mu}_{\,;\mu}$ the expansion scalar, $\sigma_{\mu\nu}=\theta_{\mu\nu}-\frac{1}{3}h_{\mu\nu}\theta$ the shear tensor, $\zeta$ the bulk viscosity, $\eta$ the shear viscosity, and $Q^{\mu}=-\kappa h^{\mu\nu}(T_{,\nu}+TU^{\nu}U_{\mu;\nu})$ the heat flux density four-vector with $\kappa$ the thermal conductivity.

In the case of thermal equilibrium, $Q_{\mu}=0$. Moreover,the term of shear viscosity vanished when a completely isotropic
unverse is assumed. In the isotopic and homogeneous Friedmann-Robertson-Walker (FRW) metric,

\begin{equation}\label{eqn:frw}
ds^2=-dt^2+a^2(t)\left(\frac{dr^2}{1-kr^2}+r^2d\Omega^2\right),
\end{equation}
where $k=-1,0,1$ being the curvature parameter, Eq.~\eqref{equ:Tmunu} can be rewritten as
\begin{equation}
T_{\mu\nu}=\rho U_{\mu}U_{\nu}+(p-\theta\zeta)h_{\mu\nu}.
\end{equation}
Here $\theta=3H$ where $H=\frac{\dot{a}}{a}$ is the Hubble parameter and the dot denotes differential with respect to cosmic time $t$.
Therefore the corresponding Friedmann equation, considering the case of flat space-time, is of the form
\begin{align}\label{eqn:fried}
H^2 &= \frac{8\pi G}{3}\rho, \\
\dot{H}+H^2 &= \frac{4\pi G}{3}\left(\rho+3\tilde{p}\right),
\end{align}
as well as the equation of energy conservation for a complete dynamics system,
\begin{equation}\label{eqn:continu}
\dot{\rho}+\left(\rho+\tilde{p}\right)\theta=0,
\end{equation}
where $\tilde{p}=p-\theta\zeta$ is the effective pressure.
If we assume the matter presented is cold thereby pressureless, $\tilde{p}$ will have a simple form of $- \theta\zeta$. So far we can principally solve the matter density $\rho$ with respect to cosmic time $t$ when the explicit form of $\zeta$ is specified.

Combining equations \eqref{eqn:fried}, \eqref{eqn:continu} and the current value of Hubble parameter $H_0$, we obtain the following dimensionless equation,
\begin{equation}\label{eqn:visco2}
\frac{\dot{h}}{H_0}+\frac{3}{2}h^2=\xi h,
\end{equation}
in which $h\equiv\frac{H}{H_0}$ and $\xi\equiv\frac{2}{3}\frac{H_0\zeta}{8\pi G}$ are the dimensionless hubble parameter and viscosity, respectively. Using the relation
\begin{equation}
\frac{\mathrm{d}}{\mathrm{d}t}=\frac{\dot{a}}{a}\frac{\mathrm{d}}{\mathrm{d}\ln a}
\end{equation}
and let $'$ denote the differential respect to conformal time $\ln a$, we obtain a dimensionless equation,
\begin{equation}\label{eqn:visco}
h'+\frac{3}{2}h=\xi.
\end{equation}
\subsection{Model \uppercase\expandafter{\romannumeral 1}: Viscosity with Decelerating Parameter $q$}
In this subsection we consider the model by taking into account of the decelerating parameter $q$. The bulk viscosity reads,
\begin{equation}\label{eqn:bulkviscosity}
\zeta=\zeta_{0}+(\zeta_{1}-\zeta_{2}q)H,
\end{equation}
in which $q\equiv-\frac{\ddot{a}}{aH^2}$ is referred as the decelerating parameter, and $\zeta_1$,$\zeta_2$,$\zeta_3$ are constant, respectively. By introducing decelerating parameter $q$ into the viscosity, the model now can take account of the role played by the transition of cosmic evolution phases.

By transformation $\zeta_0=\frac{12\pi G}{H_0}\xi_0$, $\zeta_1=\frac{12\pi G}{H_0^2}\left(\xi_1-\xi_2\right)$ and $\zeta_2=\frac{12\pi G}{H_0^2}\xi_2$, we obtain the dimensionless form of viscosity,
\begin{equation}
\xi=\xi_0+\xi_1h+\xi_2h',
\end{equation}
where the prime denotes $\frac{\mathrm{d}}{\mathrm{d}\ln a}$. Considering the condition $h(a_0)=1$, Eq.~\eqref{eqn:visco} can be solved analytically when $\xi_1\neq\frac{3}{2}$ and $\xi_2\neq 1$ as
\begin{equation}\label{eqn:solutionq}
h(a)= \frac{2\xi_0}{3-2\xi_1}+\left(1-\frac{2\xi_0}{3-2\xi_1}\right)\left(a/a_0\right)^{-\frac{3-2\xi_1}{2-2\xi_2}}.
\end{equation}
After integration we obtain the corresponding scale factor,
\begin{align}
a(t)=a_0\bigg(\frac{3-2\xi_1}{2\xi_0}&e^{\frac{\xi_0}{1-\xi_2}H_0\left(t-t_0\right)}-\frac{3-2\xi_1}{2\xi_0}+1\bigg)^{\frac{2-2\xi_2}{3-2\xi_1}}.
\nonumber\\
&
\end{align}
Hence the Hubble parameter with respect to $t$ reads
\begin{equation}\label{eqn:Ht}
H\left(t\right)=H_0\frac{2\xi_0}{3-2\xi_1}\frac{e^{\frac{\xi_0}{1-\xi_2}H_0\left(t-t_0\right)}}
{e^{\frac{\xi_0}{1-\xi_2}H_0\left(t-t_0\right)}-\frac{3-2\xi_0-2\xi_1}{3-2\xi_1}},
\end{equation}
and the decelerating parameter $q$ is
\begin{equation}\label{eqn:qt}
q=-1+\left(\frac{3-2\xi_0-2\xi_1}{2-2\xi_2}\right)e^{-\frac{\xi_0}{1-\xi2}H_0\left(t-t_0\right)}.
\end{equation}
Substitute Eq.~\eqref{eqn:Ht}, \eqref{eqn:qt} to Eq.~\eqref{eqn:bulkviscosity} we obtain the evolution of bulk-viscosity,
\begin{align}
\zeta(t)=\frac{12\pi G}{H_0}&\bigg(\frac{3\xi_0}{3-2\xi_1}+\frac{\xi_0(2\xi_1-3\xi_2)(3-2\xi_0-2\xi_1)}{(3-2\xi_1)^2(1-\xi_2)}
\nonumber\\
&\times\frac{1}{e^{\frac{\xi_0}{1-\xi_2}H_0(t-t_0)}-\frac{3-2\xi_0-2\xi_1}{3-2\xi_1}}\bigg).
\end{align}
\subsection{Model \uppercase\expandafter{\romannumeral 2}: Viscosity with $H^2$}
In this subsection we consider the viscosity model with Hubble parameter up to quadratic order,
\begin{equation}
\zeta=\zeta_0+\zeta_1H+\zeta_2H^2.
\end{equation}
The corresponding dimensionless viscosity is given as
\begin{equation}
\xi=\xi_0+\xi_1h+\xi_2h^2,
\end{equation}
We should note that it is another case of stronger coupling between matter and dark components since the term $H^2\sim h^2$ in proportion to velocity square (kinetic energy) takes into account of the energy transference. We will discuss the detailed feature of this model in Section~\ref{result} and ~\ref{conclusion} The solution of this model depends on the sign of $\Delta\equiv\left(\frac{3}{2}-\xi_1\right)^2-4\xi_0\xi_2$.
\subsubsection{$\Delta<0$}
In the case of $\Delta<0$, when $2\xi_1+4\xi_2-3\neq 0$ the evolution of universe is,
\begin{align}\label{eqn:solutionh2_1}
  h=\frac{3-4\xi_0-2\xi_1}{2\xi_1+4\xi_2-3}&+\left(1-\frac{3-4\xi_0-2\xi_1}{2\xi_1+4\xi_2-3}\right)
  \nonumber\\
  &\times\frac{1}{1-\frac{2\xi_1+4\xi_2-3}{2\sqrt{-\Delta}}\tan{\frac{\sqrt{-\Delta}}{2}\ln\frac{a}{a_0}}}.
\end{align}
Once more, integration is needed to obtain the behavior of scale factor with respect to $t$. However, it can not be shown explicitly due to the complexity. Therefore we will directly assign the best-fit values for the coefficients and calculate the corresponding cosmic quantities numerically.
\subsubsection{$\Delta>0$}
In the case of $\Delta>0$, an alternative expression is yielded when $2\Delta\neq 2\xi_1+4\xi_2+3$ and $\xi_2\neq 0$ ($H^2$ term appears) as
    \begin{align}\label{eqn:solutionh2_2}
    h=&\frac{\left(3-2\xi_1\right)-2\sqrt{\Delta}}{4\xi_2}+\frac{\sqrt{\Delta}}{\xi_2}
    \frac{2\sqrt{\Delta}+(2\xi_1+4\xi_2+3)}{2\sqrt{\Delta}-(2\xi_1+4\xi_2+3)}
    \nonumber\\
    &\times\left(\left(\frac{a}{a_0}\right)^{\frac{\sqrt{\Delta}}{\xi_2}}-
    \frac{2\sqrt{\Delta}+(2\xi_1+4\xi_2+3)}{2\sqrt{\Delta}-(2\xi_1+4\xi_2+3)}\right)^{-1},
    \end{align}
    for simplicity we rewrite this formula as
    \begin{equation}\label{eqn:solution2_3}
    h=\left(1-\frac{\tilde{\xi_1}\tilde{\xi_2}}{\tilde{\xi_1}-1}+\frac{\tilde{\xi_1}\tilde{\xi_2}}{\tilde{\xi_1}-(a/a_0)^{\tilde{\xi_2}}}\right)
    \end{equation}
    by redefining $\tilde{\xi_1}=\frac{2\sqrt{\Delta}+(2\xi_1+4\xi_2+3)}{2\sqrt{\Delta}-(2\xi_1+4\xi_2+3)}$ and $\tilde{\xi_2}=\frac{\sqrt{\Delta}}{\xi_2}$. From $\frac{\dot{a}}{a}=hH_0$ an expression of $a(t)$ is obtained by integration, which is
    \begin{align}\label{eqn:va2}
    &\frac{a(t)^{\tilde{\xi_1}-\tilde{\xi_1}\tilde{\xi_2}-1}}{\left(\tilde{\xi_1}-a(t)^{\tilde{\xi_2}}\right)^{\tilde{\xi_1}-1}}
    =C_1(\tilde{\xi_1}-\tilde{\xi_1}\tilde{\xi_2}-1)
    \nonumber\\
    &\times\exp\left(\frac{(\tilde{\xi_1}-\tilde{\xi_1}\tilde{\xi_2}-1)
    (\tilde{\xi_1}-\tilde{\xi_2}-1)}{\tilde{\xi_1}-1}H_0(t-t_0)\right),
    \end{align}
    where $C_1$ is an integrating constant determined by the condition $a(t_0)=a_0$. We cannot solve $a(t)$ explicitly due to the complex form. Nevertheless, we can study the future behavior by appropriate approximation according to the numerically fitting result.
    When $a(t)$ grows with time flying so that $|a(t)^{\tilde{\xi_2}}|\gg\tilde{\xi_1}$, Eq.~\eqref{eqn:va2} can be easily calculated as
    \begin{align}
    a(t)\simeq& a_0
    \exp\left(\frac{(\tilde{\xi_1}-\tilde{\xi_1}\tilde{\xi_2}-1)
    (\tilde{\xi_1}-\tilde{\xi_2}-1)}{(\tilde{\xi_1}-1)(\tilde{\xi_1}+\tilde{\xi_2}-2\tilde{\xi_1}\tilde{\xi_2})}H_0(t-t_0)\right)
    \nonumber\\
    &
    \end{align}
    Under the same approximation the Hubble parameter with respect to $t$ is constant,
    \begin{align}
    H(t)\simeq& H_{late-time}=\frac{(\tilde{\xi_1}-\tilde{\xi_1}\tilde{\xi_2}-1)
    (\tilde{\xi_1}-\tilde{\xi_2}-1)}{(\tilde{\xi_1}-1)(\tilde{\xi_1}+\tilde{\xi_2}-2\tilde{\xi_1}\tilde{\xi_2})}H_0
     \nonumber\\
    &
    \end{align}
\section{Data Analysis}\label{fit}
\subsection{Type Ia Supernovae}
\begin{figure}[htbp]
\centering
\subfigure[The SNe Ia dataset contains 557 points range from redshift 0.015 to 1.4.]{\label{fig:sne}
\includegraphics[width=0.8\linewidth]{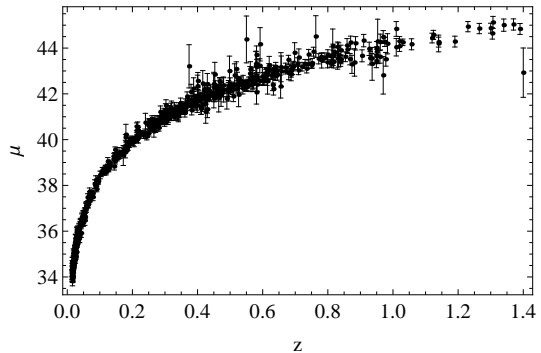}}
\\
\subfigure[The OHD dataset constains 18 points range from redshift 0.09 to 1.75.]{\label{fig:hz}
\includegraphics[width=0.8\linewidth]{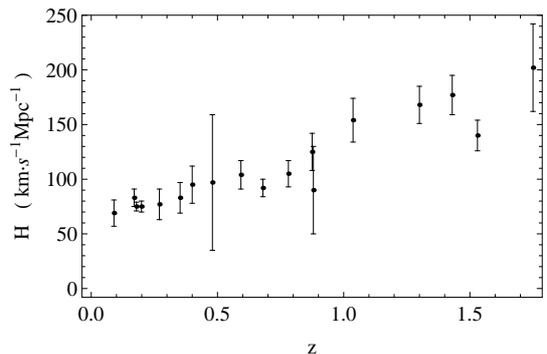}}
\caption{The observed SNe Ia and OHD data (with error bar) that we employ for best-fitting approach.}
\end{figure}

With its extraordinary property of uniform absolute magnitude, observation of Type Ia Supernovae suggests a way for inspecting the history of universe by constructing the relation between the red-shift $z$ and luminosity distance. It is believed that the SNe Type Ia observation can provide the most direct evidence of cosmic accelerating expansion. In this paper we use the Union2 SNe Ia dataset~\cite{union2} for best-fitting, which compiles 557 SNe Ia covering the redshift range $z =[0.015, 1.4]$, as shown in Fig. \ref{fig:sne}. It extends the Union dataset by including new data points at low and intermediate redshifts discovered by the CfA3 and SDSS-II Supernova Search projects, as well as 6 new SNe discovered by the Hubble Space Telescope at high z.

To perform the $\chi^2$ statistics analysis, the theoretical distance modulus is defined as
\begin{equation}
  \mu_{th}=5\log_{10}D_{L}\left(z\right)+\mu_{0},
\end{equation}
where $D_{L}\equiv H_{0}d_(z)$ is the dimensionless luminosity and $d_{L}=\left(1+z\right)d_{M}\left(z\right)$, with $d_{M}\left(z\right)$ denoting the co-moving distance,
\begin{equation}
d_{M}=\int_{0}^{z}\frac{1}{H(z')}\mathrm{d}z'.
\end{equation}
Therefore, the corresponding $\chi^{2}_{SNe}$ function is calculated from
\begin{equation}
\chi^{2}_{SNe}=\sum^n_{i=1}\left[\frac{\mu_{obs}(z_{i})-\mu_{th}(z_{i};\mathbf{\vartheta};\mu_0)}{\sigma_{obs}(z_i)}\right]^2,
\end{equation}
which should be minimized consequently with properly choosing of $\mu_0$ and model parameters $\mathbf{\vartheta}$.
The minimization with respect to $\mu_0$ can be made trivially (Ref.~\cite{sniatrick}) by expanding $\chi^{2}_{SNe}$ as
\begin{equation}\label{eqn:sneiamu}
\chi^{2}_{SNe}=A-2\mu_0B+\mu^2_0C,
\end{equation}
where
\begin{align}
A(\mathbf{\vartheta})&=\sum^n_{i=1}\left[\frac{\mu_{obs}(z_{i})-\mu_{th}(z_{i};\mathbf{\vartheta};\mu_0=0)}{\sigma_{obs}(z_i)}\right]^2, \\
B(\mathbf{\vartheta})&=\sum^n_{i=1}\frac{\mu_{obs}(z_{i})-\mu_{th}(z_{i};\mathbf{\vartheta};\mu_0=0)}{\sigma_{obs}^2(z_i)}, \\
C(\mathbf{\vartheta})&=\sum^n_{i=1}\frac{1}{\sigma_{obs}^2(z_i)}.
\end{align}
Thus $\mu_0$ is automatically minimized as $\mu_0=\frac{B}{C}$ by calculating the following transformed $\chi^2$:
\begin{equation}
\tilde{\chi}^2_{SNe}(\vartheta_{\alpha})=A(\mathbf{\vartheta})-\frac{B^2(\mathbf{\vartheta})}{C}.
\end{equation}
\subsection{Observational Hubble Parameter Data}
Recently, the direct measurement of $H(z)$ arouses much attention. In this paper we use the OHD dataset of 18 measurement points collected in Ref.~\cite{ohd}, which combines the observational constraints on the Hubble parameter available so far. It includes 8 points of the
Hubble parameter in the range $0.1<z<1.75$ from the relative dating of 32 passively evolving
galaxies~\cite{ohd1}, 2 more observations of the Hubble parameter at $z\sim0.5$ and
$z\sim0.9$ obtained by high-quality spectra of red-envelope galaxies in 24 galaxy clusters~\cite{ohd2}, and 8 new high-accuracy estimates of H(z) provided by Ref.~\cite{ohd3} for helpful compiling, extracted from several spectroscopic surveys among a large sample of 11324 early type galaxies.
All the values are reported in Table~\ref{Htab} in the appendix and shown in Fig.~\ref{fig:hz}.

The $\chi^2$ for Observational Hubble Data is
\begin{equation}
\chi_{OHD}^2=\sum_{i=0}^{n}\frac{[H_{0}h(z_{i})-H_{obs}(z_{i})]^2}{\sigma_{i}^2},
\label{chi_OHD}
\end{equation}
where $h(z)\equiv H(z)/H_{0}$ is the dimensionless Hubble parameter. Following the same approach as $\chi^2_{SNe}$, the minimization with respect to $H_0$ can be satisfied automatically by introducing the transformed $\chi^2$ function:
\begin{equation}
\tilde{\chi}^2_{OHD}=-\frac{B^2}{A}+C,
\end{equation}
where
\begin{align}
A&=\sum_{i=1}^{n}\frac{h^2(z_i)}{\sigma_i^2},
\\
B&=\sum_{i=1}^{n}\frac{h(z_i)H_{obs}(z_i)}{\sigma_i^2},
\\
C&=\sum_{i=1}^{m}\frac{H_{obs}^2(z_i)}{\sigma_i^2}.
\end{align}
\section{Results}\label{result}
\begin{figure}[!htbp]
\centering
\subfigure[The $a-t$ relation diagram. Basically there are three phases during the evolution: An exponentially inflationary scenario at the very beginning, followed by a decelerating phase (matter dominant era), eventually enters into the accelerating expansion (dark energy dominant era). We can see the results correspond to $\Lambda$CDM model very well and constraint by SNe performs better than OHD.]{\label{fig:M1a}
\includegraphics[width=0.8\linewidth]{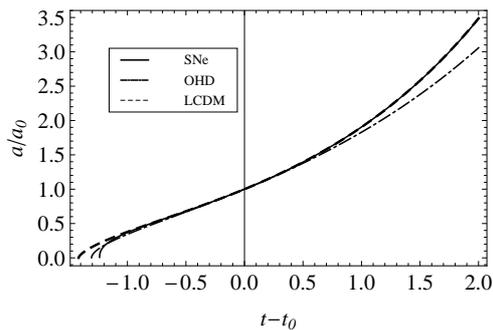}}
\\
\subfigure[The $H-t$ relation diagram. The curves of SNe and $\Lambda$CDM still stay close while the result of OHD deviates slightly.]{\label{fig:M1h}
\includegraphics[width=0.8\linewidth]{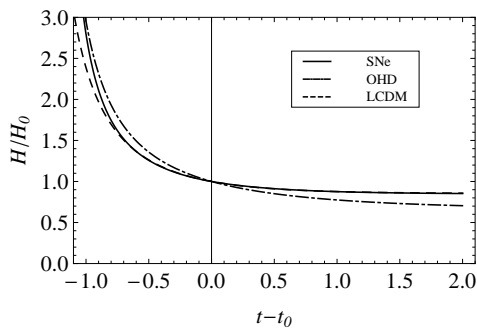}}
\\
\subfigure[The $q-t$ relation diagram. In this panel we see significant distinction between the proposed model and $\Lambda$CDM in the early time of universe. Unlike the initial value 0.5 in $\Lambda$CDM, the decelerating parameter of this viscous model starts at a relatively large value and rapidly falls to and crosses over zero, followed by a smooth evolution to -1.]{\label{fig:M1q}
\includegraphics[width=0.8\linewidth]{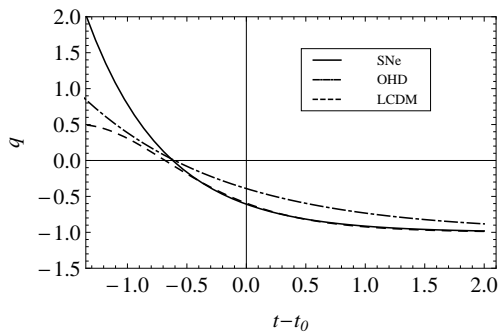}}
\caption{Resulting evolution of cosmic quantities of Model I when best-fit values assigned. Result of $\Lambda$CDM model is also plotted for comparison.}\label{fig:M1}
\end{figure}

\begin{figure}[htbp]
\centering
\subfigure[The $a-t$ relation diagram. We find no big bang in this case, instead of which the evolution of scale factor performs as two de Sitter-like expansion combined via a transition phase. However, in region of observational data, the model corresponds to $\Lambda$CDM very well.]{\label{fig:M3aD}
\includegraphics[width=0.8\linewidth]{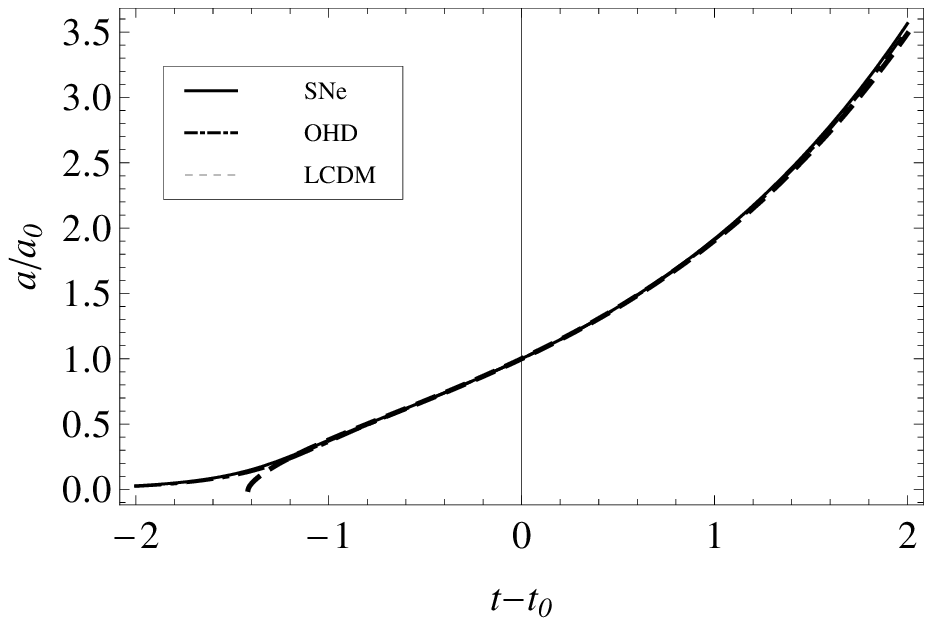}}
\\
\subfigure[The $H-t$ relation diagram. We can see clearly the three phases of the evolution in this panel, beginning with a relatively larger value of $H$, then going down smoothly during the transitionary era, and ended up with a smaller value of $H$.]{\label{fig:M3H}
\includegraphics[width=0.8\linewidth]{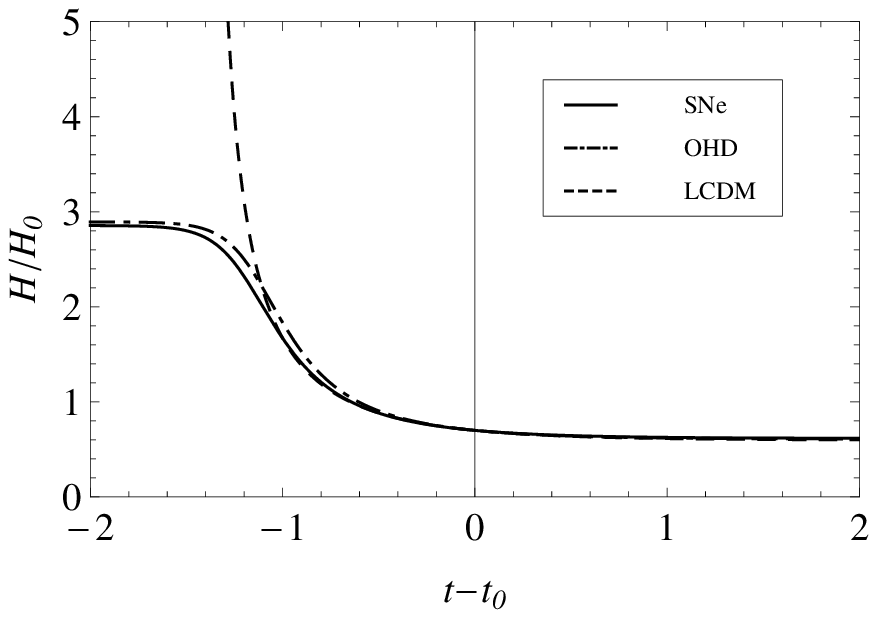}}
\\
\subfigure[The $q-t$ relation diagram. In accord with the three phases, the decelerating parameter $q$ goes from negative to positive in the pre-big bang era, then runs very closely to that of $\Lambda$CDM model in the data region and future.]{\label{fig:M3q}
\includegraphics[width=0.8\linewidth]{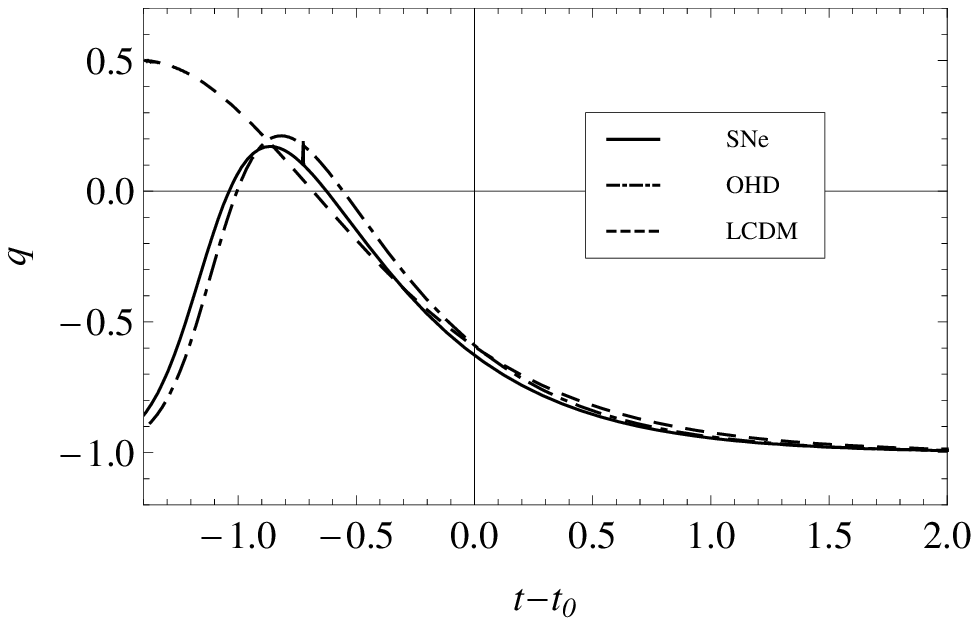}}
\caption{Resulting evolution of cosmic quantities of Model II when $\Delta>0$. In this case we obtain a universe that has no big bang. It acts as de Sitter expansion with different values of $H$ in both the very past and far future. The range of data fitting locates in the transitionary era between the two phases, which also mimics precisely the behavior of $\Lambda$CDM in that region when the best-fit values are assigned.}\label{fig:M3}
\end{figure}

In this section we perform best-fitting to the models mentioned in Section \ref{solution} and plot the evolution of relative quantities, respectively.
For computing feasibility, we shall rewrite Eq. \eqref{eqn:solutionq}, \eqref{eqn:solutionh2_1}, and \eqref{eqn:solutionh2_2} with respect to redshift $z$ and introduce alternative parameters for smoothness of the gradient calculated in minimizing $\chi^2$.
\begin{itemize}
  \item Model I, bulk viscosity with $q$,
  \begin{equation}
  h(z)=(1-\lambda_1)+\lambda_1(1+z)^{\lambda_2},
  \end{equation}
  with $\lambda_1=1-\frac{2\xi_0}{3-2\xi_1}$ and $\lambda_2=\frac{3-2\xi_1}{2-2\xi_2}$.
  \item Model II.a, bulk viscosity with $H^2$, $\Delta<0$,
\begin{equation}
  h(z)=\lambda_1+(1-\lambda_1)\frac{1}{1-\lambda_2\tan(\lambda_3\ln(1+z))},
\end{equation}
  with $\lambda_1=\frac{3-4\xi_0-2\xi_1}{2\xi_1+4\xi_2-3}$, $\lambda_2=\frac{2\xi_1+4\xi_2-3}{2\sqrt{-\Delta}}$ and $\lambda_3=\frac{\sqrt{-\Delta}}{2}$.
  \item Model II.b, bulk viscosity with $H^2$, $\Delta>0$,
 \begin{equation}
  h(z)=1-\frac{\lambda_1\lambda_2}{\lambda_1+1}+\frac{\lambda_1\lambda_2}{(1+z)^{-\lambda_2}+\lambda_1},
 \end{equation}
 with the $\tilde{\xi}$ in Eq.~\eqref{eqn:solution2_3} replaced by $\lambda$.
\end{itemize}
The results of data fitting is listed in Table~\ref{fitresult}, while the calculated evolution of cosmic quantities plotted respectively in Fig. ~\ref{fig:M1},~\ref{fig:M2} and~\ref{fig:M3}. For comparison, we also plot the evolution of $\Lambda$CDM model, $H^2=\Omega_m a^{-3}+1-\Omega_m$, with $\Omega_m=0.27$.
\begin{figure}[!ht]
\centering
\subfigure[Model I. A transition from negative to positive viscosity is shown in this figure. The negative viscosity corresponds to the era of inflation which has sufficiently large $q$. Then the value turns positive and increases until approaches to a constant value in the future.]{\label{fig:M1v}
\includegraphics[width=0.8\linewidth]{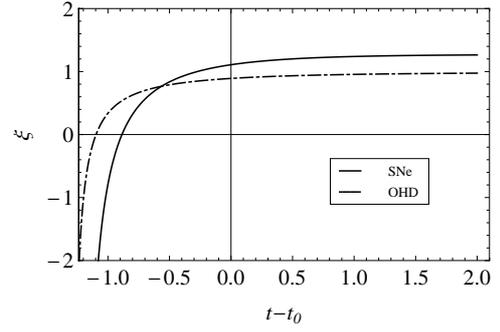}}
\\
\subfigure[Model II with $\Delta<0$. It increases from negative to positive and ends up with zero after a summit at late-time.]{\label{fig:M2v}
\includegraphics[width=0.8\linewidth]{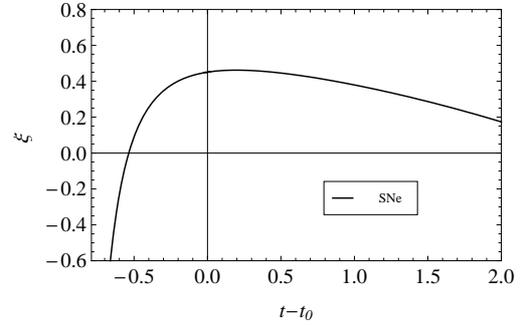}}
\\
\subfigure[Model II with $\Delta>0$. It begins at a positive value then drops below zero and recovers asymptotically to a lower positive value. The negative viscosity corresponds to the deceleration of mater dominant phase.]{\label{fig:M3v}
\includegraphics[width=0.8\linewidth]{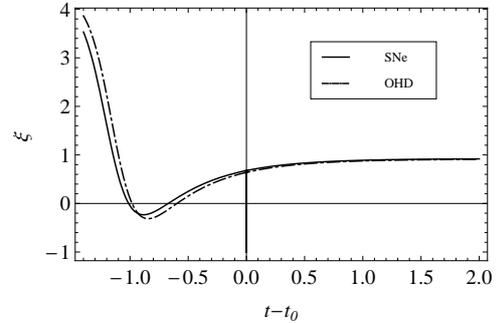}}
\caption{The $\xi-t$ relation diagram of Model I, II.a and II.b when best-fit values of parameters are given.}\label{fig:Mv}
\end{figure}
\begin{figure}[htbp]
\centering
\includegraphics[width=0.8\linewidth]{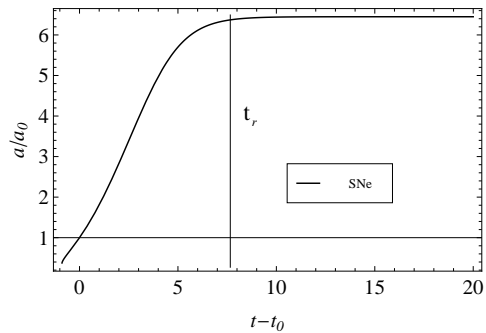}
\caption{The far-future $a-t$ relation diagram of Model II with $\Delta <0$. The scale factor in this case behaves exotically that it climbs up to an upper bound at time $t_c$, then remains constant. We will discuss the feature in Section~\ref{conclusion}.}\label{fig:M2aAll}
\end{figure}
\begin{table}[htbp]
\centering
\begin{tabular}{cccc}
\hline\hline
        &~~~I~~~&~II, $\Delta<0$~&~II, $\Delta>0$~\\
\hline
        &  $\lambda_1=0.15312$  &$\lambda_1=0.11577$  &$\lambda_1=0.039259$\\
SNe     &  $\lambda_2=2.5566$   &$\lambda_2=0.76136$  &$\lambda_2=3.2000$\\
        &  $\chi^2_{min}=542.38$&$\lambda_3=0.88993$  &$\chi^2_{min}=542.59$\\
        &                       &$\chi^2_{min}=542.15$&\\
\hline
        &  $\lambda_1=0.34146$  & &$\lambda_1=0.041795$ \\
OHD     &  $\lambda_2=1.7833$   & N/A &$\lambda_2=3.2641$ \\
        &  $\chi^2_{min}=12.513$& &$\chi^2_{min}=12.005$  \\
\hline\hline
\end{tabular}
\caption{Best-fit results of Model I, II.a, and II.b}\label{fitresult}
\end{table}
\begin{figure}[!htp]
\centering
\subfigure[The $a-t$ relation diagram. We can see in this case the evolution of scale factor coordinates with $\Lambda$CDM model in the range of observation data, and the evolution curve has a similar trend with Model I (Big bang, inflation, deceleration, and acceleration). Nevertheless, there are two main differences: The age of universe is much shorter than that predicted in $\Lambda$CDM; The future behavior is widely divergent (see Fig.~\ref{fig:M2aAll}).]{\label{fig:M2aD}
\includegraphics[width=0.8\linewidth]{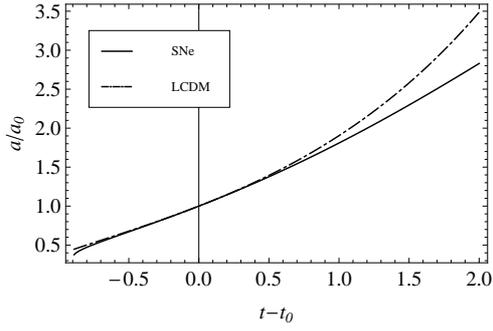}}
\\
\subfigure[The $H-t$ relation diagram. Instead of a monotonic descending followed by an asymptotical constant value as suggested by $\Lambda$CDM model, the evolution of Hubble parameter in $H^2$ model decreases more dramatically at early time, and goes towards zero in the future.]{\label{fig:M2h}
\includegraphics[width=0.8\linewidth]{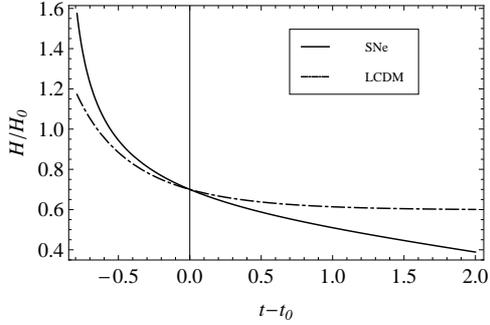}}
\\
\subfigure[The $q-t$ relation diagram. In accord with the future halt of expansion, the run of $q$ is essentially different from that of $\Lambda$CDM.]{\label{fig:M2q}
\includegraphics[width=0.8\linewidth]{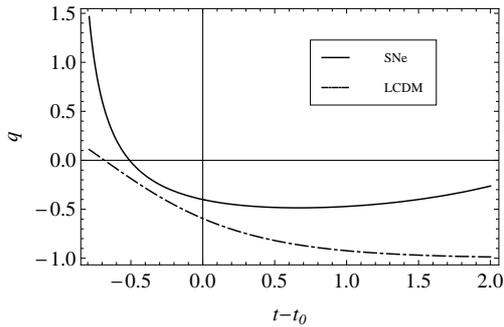}}
\caption{Resulting evolution of cosmic quantities of Model II with best-fit values assigned when $\Delta<0$. We only plot that constrained by SNe data, because a poor confidence of estimated parameter values is confronted when OHD dataset is employed.}\label{fig:M2}
\end{figure}

\section{Conclusions and Discussions}\label{conclusion}
In this paper we continue our previous work on bulk viscous (dark
fluid) cosmology by focusing on two explicit viscous models and
studying the viability of accommodating the observed accelerating
expansion of the universe without introducing the mysterious dark
energy. We employ the Union2 data and OHD dataset to generate the
best-fit values for the models, and then accordingly draw the
corresponding evolution diagrams, respectively. As comparison, we also demonstrate that the resulting evolutions of such models are precisely consistent with $\Lambda$CDM model in the region of observational data.
Both of our models are proposed in the context of assumed tight coupling between the
cosmic components, and by arousing a time-varied bulk viscosity as
constructed, we see the cosmic evolutions of such models can be
greatly different from that suggested by the $\Lambda$CDM.

We present our main conclusions and discussions in order as
following. \textbf{Model I}, $\zeta=\zeta(q,H)$, mimics the universe
evolution of $\Lambda$CDM model perfectly. We can see from
Fig.~\ref{fig:M1a} that the unverse has an exponentially inflation
after the Big Bang, and soon enters into deceleration when in matter
dominant era, followed by a late-time accelerating expansion. The
monotonic increasing bulk viscosity (Fig.~\ref{fig:M1v}) with
respect to both $q$ and $H$ provides a connection among the
pressureless matter therefore behaves as effective dark energy in
the late-time evolution. We should notice that although the age of
universe is smaller in this model than that of $\Lambda$CDM,
it shares the most similarities with those of the $\Lambda$CDM
model, which makes this kind of bulk viscosity component model a
successful substitution for dark energy.

\textbf{Model II}, $\zeta=\zeta(H,H^2)$ as another case of viscous
cosmology model, suggests two alternative evolutions due to
different values of parameters. In the \textbf{Case of $\Delta<0$},
we obtain a bound universe that the expansion ends at far future
$t=t_c$, when $H$ and $\dot{H}$ both vanishes (static then)
(Fig.~\ref{fig:M2aAll}). However, it fits the data quite well in
corresponding data region as seen in Fig.~\ref{fig:M2aD}. The halt
confronted in the far future is the result of the viscosity with
$H^2$ that drives \eqref{eqn:visco2} to a trivial solution.
From the numeric calculation we estimate the value of that time,
$t_c-t_0\simeq70.2\mathrm{Gyr}$. In the \textbf{Case of $\Delta>0$},
a no-big-bang unverse is obtained. It performs as a de-Sitter
expansion in both the very past and very future with different
values of $H$. The range of data fitting locates in the
transitionary phases between the two eras. In the region it mimics
the observed acceleration successfully. From Fig.~\ref{fig:M3H} and
Fig.~\ref{fig:M3v} we see that the values of $H$, $q$ and $\xi$ go
to approximated constants when $t_c-t_0\simeq19.4\mathrm{Gyr}$.

We are able to investigate the universe fate of our cosmology models
by means of the rip or singularity. Singularity behavior is
classified in Ref.~\cite{singupro} as follows:
\begin{itemize}
   \item Type I (``Big Rip"): For $t\rightarrow t_s$, $a\rightarrow\infty$, $\rho\rightarrow\infty$, and $|p|\rightarrow\infty$;
   \item Type II (``Sudden"): For $t\rightarrow t_s$, $a\rightarrow a_s$, $\rho\rightarrow\rho_s$, and $|p|\rightarrow\infty$;
   \item Type III: For $t\rightarrow t_s$, $a\rightarrow a_s$, $\rho\rightarrow\infty$, and $|p|\rightarrow\infty$;
   \item Type IV: For $t\rightarrow t_s$, $a\rightarrow a_s$, $\rho\rightarrow 0$,and $|p|\rightarrow 0$, and higher derivatives of H diverge.
 \end{itemize}
In our cases, an effective pressure is arisen as a result of the
presence of bulk viscosity, reads $\tilde{p}=-3H\zeta$. Hence the
evolution of $|p|$ is similar to that of bulk viscosity in the
future because of the asymptotic constant for Hubble parameter.
According to Fig.~\ref{fig:Mv},~\ref{fig:M1},~\ref{fig:M2} and
~\ref{fig:M3}, none of the finite singularity is confronted,
therefore viscosity in our model as a practical substitution of dark
energy candidate is better than some other theories (e.g. phantom
dark energy).

The Rip behavior is another aspect we interest in. Three types of
Rips are discussed in Ref.~\cite{pseudorip} according to the
evolution of inertial force on a mass $m$ as seen by a gravitational
source separated by a co-moving distance $l$, defined as
$F_{inert}=ml(\dot{H}+H^2)$.
\begin{itemize}
   \item ``Big Rip": For $t\rightarrow t_s$, $F_{inert}\rightarrow \infty$;
   \item ``Little Rip": For $t\rightarrow \infty$, $F_{inert}\rightarrow \infty$;
   \item ``Pseudo Rip": For $t\rightarrow \infty$, $F_{inert}$ never goes infinite but has an upper bound.
 \end{itemize}
Rips occur when the bound structures are disintegrated by the cosmic
expansion, thereby being an important feature in testifying models.
In our models, the inertial force is calculated by $F_{inert}\propto
H^2+9H\zeta$, which never varies towards infinite during the
evolution, i.e., they are free of little rip. What's more, although
in model I and II.a there are maximums of $F_{inert}$, they are
merely fail pseudo rips because of the small value that cannot
disassemble the bound structure of the universe, i.e. they are free
of pseudo rip.

Our real universe is filled with viscosity media, therefore
cosmology based on imperfect fluid is proposed reasonably, which is
proved to be viable in interpreting the observed cosmic accelerating
expansion. We see that for the theories we study in this paper, the
presence of bulk viscosity allows for a much more flexible evolution
destination, when avoiding any type of singularity or rip. On the
other hand, in this work we have only pressureless (cold dark)
matter and an effective pressure provided by viscosity term,
therefore it is hard to recognize $\Omega_m$ from the total matter
density, as they are coupled tightly, if the form of viscosity or
its evolution behavior is complex. This explains why we discard to
use CMB-shift and BAO data for joint constraint analysis. We also
show that, as a promising probe for cosmological fitting,
observational Hubble data performs very well and gives reliable
results. It is believed that with much more high-precision data
points released or published, the OHD dataset will play an even
important role as an independent and direct measurement.

Instead of adding bulk viscosity to $T_{\mu\nu}$, the R.H.S. of
Einstein equation, modification on the gravity theory, the left side
of Einstein equation or the gravity sector, is another approach of
accommodating the cosmic late-time acceleration expansion. It is
possible to be reconstructed effectively from the left side to the
right side of the equation, which is also intriguing and we will be
present that explorations elsewhere. It is also expected in the
future work to give additional constraints on the parameters of
viscosity models from cosmic fluid dynamics.
\section*{Acknowledgement}
We are benefit from interesting discussions with Prof. S. D. Odintsov and this work is partly supported by Natural
Science Foundation of China under Grant Nos.11075078 and 10675062
and by the project of knowledge Innovation Program (PKIP) of Chinese
Academy of Sciences (CAS) under the grant No. KJCX2.YW.W10 through
the KITPC where we have initiated this present work.
\section*{Appendix}
The list of 18 OHD data used in this paper.
\begin{table}[htbp]
\centering
\begin{tabular}{ccc}
\hline\hline
        $~~~z~~~$  & $~~~H(z)\pm 1\sigma~~~$&~~~Ref.~~~\\
                         &~~km s$^{-1}$Mpc$^{-1}$&\\
\hline
0.090 &$69\pm12$ & \cite{ohd1}\\
0.170& $83\pm8$ & \cite{ohd1} \\
0.179& $75\pm4$ &\cite{ohd3}\\
0.199& $75\pm5$ &\cite{ohd3}\\
0.270& $77\pm14$ & \cite{ohd1} \\
0.352& $83\pm14$ &\cite{ohd3}\\
0.400& $95\pm17$ & \cite{ohd1} \\
0.480& $97\pm62$ &\cite{ohd2}\\
0.593& $104\pm13$ &\cite{ohd3}\\
0.680& $92\pm8$ &\cite{ohd3}\\
0.781& $105\pm12$ &\cite{ohd3}\\
0.875& $125\pm17$ &\cite{ohd3}\\
0.880& $90\pm40$ & \cite{ohd2} \\
1.037& $154\pm20$ &\cite{ohd3}\\
1.300& $168\pm17$ & \cite{ohd1} \\
1.430& $177\pm18$ & \cite{ohd1} \\
1.530& $140\pm14$ & \cite{ohd1} \\
1.750& $202\pm40$ & \cite{ohd1} \\
\hline\hline
\end{tabular}
\caption{The set of available observational $H(z)$ data with their errors.}\label{Htab}
\end{table}

\end{document}